\documentclass[journal=jpclcd,
               manuscript=letter,
               ]{achemso}
\setkeys{acs}{articletitle=true,maxauthors=0}

\usepackage[version=3]{mhchem} 
\usepackage[T1]{fontenc}       
\usepackage{pstricks}

\usepackage{amsmath,amssymb}
\usepackage{graphicx}
\usepackage{bm}
\usepackage{algorithm}         
\usepackage{algpseudocode}     
\usepackage{siunitx}
\usepackage{subcaption}
\usepackage{multirow,booktabs}
\usepackage{cleveref}

\newcommand*\ex[1]{\langle{}#1{}\rangle}
\newcommand*\ket[1]{|#1{}\rangle}

\newcommand*\cd{c^{\dagger}}
\newcommand*\tr[1]{\textrm{#1}}
\newcommand*\mr[1]{\mathrm{#1}}

\newcommand*\mc[1]{\mathcal{#1}}
\newcommand*\ttt{\texttt}

\newcommand*\tPhi{\tilde{\Phi}}
\newcommand*{\wbxv}{$\omega$B97X-V}

\newcommand*{\dscf}{$\Delta$-SCF}

\author{Hong-Zhou Ye}
\author{Troy {Van Voorhis}}
\email{tvan@mit.edu}
\affiliation[Massachusetts Institute of Technology]
{Department of Chemistry, Massachusetts Institute of Technology, Cambridge, MA 02139}
\title[MP2 For Excited States]
{Self-consistent M{\o}ller-Plesset Perturbation Theory For Excited States}

\begin{document}


    \textit{Abstract} In quantum chemistry, obtaining a system's mean-field solution and incorporating electron correlation in a post Hartree-Fock (HF) manner comprise one of the standard protocols for ground-state calculations. In principle, this scheme can also describe excited states but is not widely used at present, primarily due to the difficulty of locating the mean-field excited states. With recent developments in excited-state orbital relaxation, self-consistent excited-state solutions can now be located routinely at various levels of theory. In this work, we explore the possibility of correcting HF excited states using M{\o}ller-Plesset perturbation theory to the second order. Among various PT2 variants, we find that the restricted open-shell MP2 (ROMP2) gives excitation energies comparable to the best density functional theory results, delivering $\sim 0.2$ eV mean unsigned error over a wide range of single-configuration state function excitations, at only non-iterative $O(N^5)$ computational scaling. 

    \begin{figure}
        \centering
        TOC Graphic \\
        \includegraphics[scale=1]{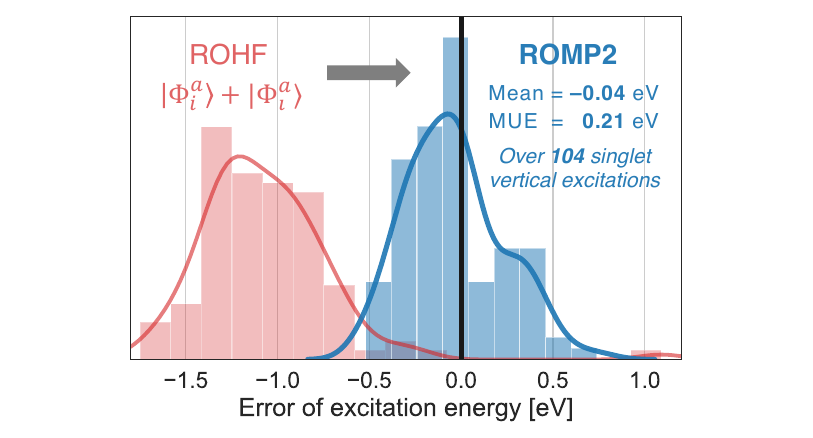}
    \end{figure}



    Recent developments in electronic structure theory have led to a revival of state-specific self-consistent methods for electronic excited states at various levels of theory, including Hartree-Fock \cite{Hiscock14JCTC,Burton16JCTC,Ye17JCP,Ye19JCTC,Michelitsch19JCP,Hait20JCTC,Hait20JCTC_II,Levi20FD,Mihalka20JMC} (HF),
    configuration interaction with singles \cite{Shea18JCP,Shea20JCTC,Zhao20JCTC,Zhao20JCP} (CIS),
    complete active space \cite{Tran19JCTC} (CAS),
    Kohn-Sham density functional theory \cite{Cheng08JCP,Evangelista13JPCA,Derricotte15PCCP,Ramos18JCP,Roychoudhury20SR,CarterFenk20JCTC} (KS-DFT),
    and variational Monte-Carlo \cite{Zhao16JCTC,Shea17JCTC,Zhao19PRL} (VMC), to name a few.
    Lying at the core of these methods is approximating an excited-state wave function by one or the linear combination of a small number of Slater determinants whose constituent orbitals are optimized self-consistently.
    This scheme parallels the ground-state HF \cite{Roothaan51RMP,Szabo89book} and KS-DFT \cite{Hohenberg64PR,Kohn65PR}, but contrasts with the linear response (LR) approach \cite{Dreuw05CR,Hirata08Inbook} where excited states are described indirectly by the response of the corresponding ground state to time-dependent electric fields.
    Due to the state-specific orbital relaxation, the self-consistent approach provides a reasonable description to excitations of charge-transfer, Rydberg, or doubly excited character at the mean-field level, \cite{Cheng08JCP,Ye17JCP,Ye19JCTC,Liu17JCTC,Tran19JCTC,Zhao20JCTC,Shea18JCP,Shea20JCTC,Hait20JCTC,Roychoudhury20SR,CarterFenk20JCTC}
    which are known to be challenging for the LR methods. \cite{Dreuw04JACS,Cheng08JCP,Ziegler10JCP,Elliott11CP,Baerends13PCCP,vanMeer14JCTC,Krykunov14JCP,Seidu15JPCA,Maitra16JCP,Hait19PCCP}
    In particular, KS-DFT-based self-consistent approaches like $\Delta$ self-consistent field \cite{Jones89RMP,Hellman04JCP,Gavnholt08PRB} (\dscf{})
    and restricted open-shell KS \cite{Filatov99CPL,Kowalczyk13JCP} (ROKS) have demonstrated quite good accuracy for excited states dominated by a single configuration state function (CSF), \cite{Kowalczyk13JCP,Hait16JCTC,Liu17JCTC,Hait20JCTC,Hait20JCTC_II,CarterFenk20JCTC}
    and the functional-dependence is mitigated compared with the LR time-dependent DFT (TDDFT). \cite{Kowalczyk13JCP,Zhao20JCTC,Hait20JCTC_II}

    A natural question is how to improve the mean-field self-consistent approach \textit{ab initio} by using electron correlation methods. Iterative methods such as coupled cluster \cite{Bartlett07RMP} (CC) have been shown successful in correcting a small set of doubly excited \dscf{} states, \cite{Lee19JCP} but can have convergence issues for excited states in general. Thus, the non-iterative M{\o}ller-Plesset perturbation theory \cite{Moller34PR} (MPPT) provides an easier route. The accuracy of MPPT depends on the quality of the reference state. \cite{Nobes91CPL,Soydas15JCTC}
    For excited states, one of the main concerns is spin-symmetry \cite{Ye19JCTC}: a spin-contaminated reference [e.g.,\ spin-unrestricted HF \cite{Szabo89book} (UHF)] can significantly deteriorate the performance of lower-order MPPT calculations. \cite{Schlegel86JCP,Gill88JCP,Kurlancheek09MP,Hait18JCTC,Hait18PCCP}
    Indeed, previous works that apply MPPT to self-consistent excited states, such as the non-orthogonal configuration interaction MPPT to the second order formulated in either the perturb-then-diagonalize \cite{Yost13JCP,Yost16JCP,Yost18JCTC} or the diagonalize-then-perturb \cite{Burton20arXiv} manner, and Excited-state MP2 \cite{Shea18JCP,Shea20JCTC,Clune20arXiv} (ESMP2),
    avoid this issue by using reference states that involve many determinants [NOCI \cite{Thom09JCP,Sundstrom14JCP} and Excited-state Mean Field \cite{Shea18JCP,Shea20JCTC} (ESMF), respectively]. Despite their success, the use of sophisticated reference states complicates the orbital optimization process and/or the subsequent PT treatment.

    In this work, we evaluate the accuracy of PT2 corrections to various HF references that possess the correct spin-symmetry for excited states. In each case, the ground-state method is well-established, and we use a different set of orbitals optimized for an excited state. We find that ROMP2 (i.e.,\ PT2 correction applied to ROKS with the HF functional) gives excitation energies comparable to the best self-consistent KS-DFT results, delivering a mean unsigned error (MUE) of $\sim 0.2$ eV over a wide range of single-CSF excitations. For the more challenging case of the heptazine molecule, ROMP2 correctly predicts a negative singlet-triplet gap while ROKS with commonly used exchange-correlation (xc) functionals fails qualitatively. These results recommend the use of ROMP2 as an $O(N^5)$-scaling \textit{ab initio} method for excited states.


    We begin with a brief review of all the PT2 variants used in this work. We restrict the discussion here to singlet excited states, but the extension to higher spins is possible, too. Throughout this work, we use $i,j,\cdots$, $a,b,\cdots$, and $p,q,\cdots$ to index occupied, virtual, and unspecified orbitals, respectively.

    UMP2 corrects the energy of a UHF state $\ket{\Phi_{\tr{U}}}$ by
    \begin{equation}    \label{eq:UMP2_E}
        E^{(2)}_{\tr{U}}
            = \ex{\Phi_{\tr{U}} | \hat{H}-\hat{f}_{\tr{U}} |
            \Phi_{\tr{U}}^{(1)}},
    \end{equation}
    where $\hat{f}_{\tr{U}}$ is the UHF Fock operator, and
    \begin{equation}    \label{eq:UMP2_Phi_1}
        \ket{\Phi_{\tr{U}}^{(1)}}
            = \frac{1}{4} \sum_{iajb} t_{iajb} \ket{(\Phi_{\tr{U}})_{ij}^{ab}},
    \end{equation}
    with $\{t_{iajb}\}$ the MP2 amplitudes. \cite{Szabo89book}
    As mentioned above, $\ket{\Phi_{\tr{U}}}$ breaks the spin-symmetry ($\ex{\hat{S}^2}_{\tr{U}} > 0$ for singlet) for most excited states, \cite{Ye19JCTC} and we only include UMP2 here for comparison.

    Spin-projected HF \cite{Scuseria11JCP,Jimenez-Hoyos12JCP} (SPHF) restores the broken spin-symmetry of a UHF state by spin-projection,
    \begin{equation}    \label{eq:SPHF_wfxn}
        \ket{\Phi_{\tr{SP}}}
            = \hat{P} \ket{\tPhi_{\tr{U}}},
    \end{equation}
    where
    \begin{equation}    \label{eq:SP_op}
        \hat{P}
            = \frac{1}{2} \int_{0}^{\pi} \mr{d}\beta\, \sin\beta\,
            \mr{e}^{\mr{i}\beta\hat{S}_y}
    \end{equation}
    for singlet, and the tilde sign in $\ket{\tPhi_{\tr{U}}}$ indicates the orbitals being optimized for the SPHF energy, $E^{(0)}_{\tr{SP}}$ (i.e.,\ variation-after-projection \cite{Scuseria11JCP}).
    The SPMP2 energy correction is \cite{Tsuchimochi14JCP}
    \begin{equation}    \label{eq:SPMP2_E}
        E^{(2)}_{\tr{SP}}
            = \frac{\ex{\tPhi_{\tr{U}} |
            (\hat{H} - E^{(0)}_{\tr{SP}}) \hat{P} |
            \tPhi_{\tr{U}}^{(1)}}}
            {\ex{\tPhi_{\tr{U}} | \hat{P} | \tPhi_{\tr{U}}}}.
    \end{equation}
    In practice, the $\beta$-integral in the spin-projection operator is evaluated numerically on a grid, \cite{Scuseria11JCP}
    \begin{equation}    \label{eq:SP_op_on_grid}
        \hat{P}
            \approx \sum_{g=1}^{N_{\tr{grid}}} w_g \hat{R}_g,
    \end{equation}
    where a uniform grid with $N_{\tr{grid}} = 6$ proves sufficient for obtaining converged excitation energies for all the examples studied in this work.

    Taking a special two-point grid ($\beta_1 = 0$, $\beta_2 = \pi$ and $w_1 = w_2 = 1$) in \cref{eq:SP_op_on_grid} leads to the so-called half-projected HF \cite{Cox76TCA-1,Smeyers76IJQC,Ye19JCTC} (HPHF),
    \begin{equation}    \label{eq:HPHF_wfxn}
        \ket{\Phi_{\tr{HP}}}
            = \frac{1}{\sqrt{2}}(\hat{1} + \hat{\mc{C}}) \ket{\tPhi_{\tr{U}}},
    \end{equation}
    where $\hat{\mc{C}}$ flips the spin of all the spin-orbitals in $\ket{\tPhi_{\tr{U}}}$. HPHF removes the spin-contamination from mixing with triplet, septet, \textit{etc.},\ which is usually sufficient for low-lying singlet excited states. \cite{Smeyers00AQC,Ye19JCTC}
    The HPMP2 energy correction follows straightforwardly from \cref{eq:SPMP2_E,eq:SP_op_on_grid}.

    Another ansatz that maintains the correct spin-symmetry is ROHF \cite{ROHF_note}, which adopts a single-CSF wave function, \cite{Helgaker00Book}
    \begin{equation}    \label{eq:ROHF_wfxn}
        \ket{\Phi_{\tr{RO}}}
            = \frac{1}{\sqrt{2}} (\hat{1} +  \hat{\mc{C}})
            \ket{(\tPhi_{\tr{R}})_i^a}.
    \end{equation}
    $\ket{\Phi_{\tr{RO}}}$ closely resembles $\ket{\Phi_{\tr{HP}}}$ except that the orbitals here are spin-restricted, rendering $\ket{\Phi_{\tr{RO}}}$ a pure singlet. Note that both HP and RO approaches are closely related to Yamaguchi's approximate spin-projection procedure \cite{Kitagawa07CPL,Nakanishi09IJQC}, which corrects a UHF/MP2 singlet using a separate triplet calculation.
    The ROMP2 energy correction is
    \begin{equation}
        E^{(2)}_{\tr{RO}}
            = \sum_{m \in \mc{S},\mc{D}}
            \frac{|\ex{m |\hat{H}-\hat{f}_{\tr{RO}}| \Phi_{\tr{RO}}}|^2}
            {\mc{E}[(\tPhi_{\tr{R}})_i^a] - \mc{E}[m]}
    \end{equation}
    where the sum runs over all singly ($\mc{S}$) and doubly ($\mc{D}$) excited configurations from $\ket{\Phi_{\tr{RO}}}$, and $\mc{E}[m]$ is the sum of orbital energies of $\ket{m}$.

    We implement all the PT2 variants above in \ttt{frankenstein} \cite{Ye20Github}, which uses \ttt{PySCF} \cite{Sun17WIRCMS} as a backend for basic SCF modules. All the PT2 variants have the same $O(N^5)$ computational scaling (with $N$ the basis size) due to the need of transforming the electron repulsion integrals (ERIs) from the atomic orbital basis to the molecular orbital basis. The actual cost of the ERI transform relative to a RMP2 calculation is $\sim 1$ (ROMP2), $3$ (UMP2), $6$ (HPMP2), and $3N_{\tr{grid}}$ (SPHF), respectively. The core orbitals are frozen for both ground and excited states, which introduces virtually no error in the calculated excitation energies shown below.


    \begin{table}[h]
        \caption{Composition of the test set used in this work.}
        \label{tab:test_set}
        \begin{tabular}{lccc}
            \hline\hline
            Type & $N_{\tr{mol}}$ & $N_{\tr{ex}}$ & example \\
            \hline
            Saturated & $4$ & $10$ & water \\
            CH-unsaturated & $7$ & $21$ & butadiene \\
            Chromophore groups & $17$ & $54$ & acetone \\
            Aromatic rings & $4$ & $19$ & furan \\
            Total & $32$ & $104$ & \\
            \hline
        \end{tabular}
    \end{table}

    We first evaluate the accuracy of excitation energies computed using these PT2 variants via comparison to high-level theory benchmarks. A test set consisting of $104$ low-lying, singlet vertical excitations from $32$ small molecules is compiled from refs.\ \citenum{Loos18JCTC,Loos20JCTC} and summarized in \cref{tab:test_set} (see Sec.\ S1 in Supporting Information for details).
    The excitation energies obtained using selected CI \cite{Huron73JCP,Giner13CJC} or high-level equation-of-motion CC \cite{Geertsen89CPL,Watts08Inbook} (EOM-CC) in the aug-cc-pVTZ basis \cite{Dunning89JCP,Kendall92JCP,Woon93JCP} are used as benchmark (also taken from refs.\ \citenum{Loos18JCTC,Loos20JCTC}).
    The test set includes mainly single-CSF excited states that are either single-pole (i.e.,\ only one transition amplitude is significant, denoted by $1 \to 1$) or multi-pole of type $1 \to n$ ($n > 1$), but excludes the multi-pole excitations of type $m \to n$ ($m,n>1$), which are multi-CSF in nature and beyond the capability of the simple ans\"a{}tze used here. Nonetheless, we will provide specific examples of $m \to n$ excitations \textit{vide infra} and discuss potential extensions of the methods used here for describing them.

    In order to obtain the various types of HF solutions for an excited state, we start the SCF cycles with non-aufbau configurations using ground-state RHF orbitals, and then employ techniques including the maximum overlap method \cite{Gilbert08JPCA} (MOM) and the squared-gradient minimization \cite{Hait20JCTC} (SGM) to prevent the SCF iteration from converging to other states. At convergence, the transition amplitudes from the RHF ground state
    \begin{equation}    \label{eq:transition_amplitudes}
        \gamma_{i \to a}
            = \ex{\Phi_{\tr{R}}^{\tr{ground}} |
            \cd_{a} c_i | \Phi^{\tr{excited}}}
    \end{equation}
    are calculated and then compared with the EOM-CCSD results (obtained using \ttt{Psi4}\cite{Parrish17JCTC}) to verify that the protocol above had yielded the correct state.
    This protocol works well for all the excitations studied in this work, and we note that for future applications, the recently developed state-targeting strategies \cite{Ye17JCP,Shea20JCTC,CarterFenk20JCTC} can be exploited readily.

    \begin{table*}[h]
        \centering
        \caption{Mean unsigned errors of the excitation energies (unit: eV) predicted by the various PT2 variants and their HF references as discussed in the main text. EOM-CCSD results are included for comparison.}
        \label{tab:stat_PT2}
        \begin{tabular}{lccccccccc}
            \hline\hline
            & \multicolumn{2}{c}{U} & \multicolumn{2}{c}{RO} & \multicolumn{2}{c}{HP} & \multicolumn{2}{c}{SP} & \multirow{2}*{EOM-CCSD} \\
            \cmidrule(lr){2-3} \cmidrule(lr){4-5} \cmidrule(lr){6-7} \cmidrule(lr){8-9}
            & HF & PT2 & HF & PT2 & HF & PT2 & HF & PT2 & \\
            \hline
            Saturated & $1.19$ & $0.09$ & $0.99$ & $0.14$ & $0.50$ & $0.13$ & $0.30$ & $0.14$ & $0.03$ \\
            CH-unaturated & $1.20$ & $0.41$ & $0.77$ & $0.24$ & $0.47$ & $0.22$ & $0.39$ & $0.19$ & $0.08$ \\
            Chromophore groups & $1.09$ & $0.45$ & $0.70$ & $0.22$ & $0.25$ & $0.32$ & $0.44$ & $0.31$ & $0.13$ \\
            Aromatic rings & $0.92$ & $0.41$ & $0.63$ & $0.18$ & $0.34$ & $0.18$ & $0.53$ & $0.29$ & $0.12$ \\
            Total & $1.09$ & $0.40$ & $0.73$ & $0.21$ & $0.33$ & $0.26$ & $0.43$ & $0.27$ & $0.11$ \\
            \hline
        \end{tabular}
    \end{table*}

    The MUEs of all four PT2 variants and their HF references are summarized in \cref{tab:stat_PT2} (see tables S1 -- S4 for the full data). The corresponding error distributions are plotted in fig.\ S2. The accuracy of the HF reference follows the trend, HPHF $>$ SPHF $>$ ROHF $>$ UHF. From fig.\ S2, the $\sim 1$ eV error of UHF corresponds to a systematic underestimation of the excitation energy. This bias, mainly caused by spin-contamination, is corrected to different extents by the other HF methods, among which HPHF achieves the lowest MUE of $\sim 0.3$ eV, due to both error cancellation and the fact that mixing with triplet dominates the spin-contamination of low-lying singlet excitations. \cite{Smeyers00AQC,Ye19JCTC} Notably, as an $O(N^4)$-scaling method, the accuracy of HPHF is also comparable to the KS-DFT results using some good xc functionals (\textit{vide infra}).


    The PT2 corrections reduce the error for all HF references, albeit to different extents. Among them, ROMP2 is the clear winner and attains $\sim 0.2$ eV MUE over all types of excitations in the test set while requiring the least computational work.
    This performance is also comparable to other $O(N^5)$-scaling LR approaches including CIS(D) \cite{HeadGordon94CPL,HeadGordon95CPL}, second-order algebraic diagrammatic construction \cite{Dreuw15WIRCMS} [ADC(2)], and EOM-CC2 \cite{Christiansen95CPL,Hattig00JCP} (table S9).
    Interestingly, even though both HPHF and SPHF are more accurate than ROHF, the PT2 corrections bring little improvement to the former two, potentially due to a less effective error cancellation.
    On the other hand, UMP2 gives a MUE twice as high as ROMP2 except for the simple cases of saturated molecules, and has a higher number of worst cases (e.g.,\ absolute error $> 1$ eV) than other PT2 methods ($\sim 10$ for UMP2, $1$ for HPMP2, and $0$ for RO/SPMP2, as can be seen from fig.\ S2). The poor performance of UMP2 arises from its perturbative treatment of the strong coupling between the two spin-conjugated excited-state configurations (i.e.,\ $\ex{\Phi_i^a|\hat{V}|\Phi_{\bar{i}}^{\bar{a}}}$), which is already included at the mean-field level for all other PT2 methods.

    \begin{table*}[h]
        \centering
        \caption{Mean unsigned errors of the excitation energies (unit: eV) predicted by various KS-DFT methods.}
        \label{tab:stat_DFT}
        \begin{tabular}{lcccccccc}
            \hline\hline
            & \multicolumn{4}{c}{U} & \multicolumn{4}{c}{RO} \\
            \cmidrule(lr){2-5} \cmidrule(lr){6-9}
            & LDA & PBE & B3LYP & $\omega$B97X-V & LDA & PBE & B3LYP & $\omega$B97X-V \\
            \hline
            Saturated & $0.21$ & $0.51$ & $0.46$ & $0.13$ & $0.12$ & $0.35$ & $0.33$ & $0.10$ \\
            CH-unaturated & $0.73$ & $0.88$ & $0.81$ & $0.43$ & $0.54$ & $0.62$ & $0.53$ & $0.19$ \\
            Chromophore groups & $0.48$ & $0.63$ & $0.53$ & $0.35$ & $0.31$ & $0.42$ & $0.28$ & $0.19$ \\
            Aromatic rings & $0.38$ & $0.48$ & $0.45$ & $0.27$ & $0.28$ & $0.42$ & $0.30$ & $0.18$ \\
            Total & $0.48$ & $0.65$ & $0.57$ & $0.33$ & $0.33$ & $0.46$ & $0.34$ & $0.18$ \\
            \hline
        \end{tabular}
    \end{table*}

    In \cref{tab:stat_DFT}, we list the results from state-specific KS-DFT calculations at UKS (i.e.,\ \dscf{}) and ROKS levels with four different xc functionals: LDA \cite{Vosko80CJP}, PBE \cite{Perdew96PRL}, B3LYP \cite{Becke93JCP}, and \wbxv{} \cite{Mardirossian14PCCP}.
    The full data can be found in tables S5 -- S8, and the error distributions are plotted in fig.\ S3. Overall, the spin-symmetry also plays a significant role here, as ROKS improves upon UKS regardless of the xc functionals. Within each class (UKS or ROKS), however, the functional dependence is scattered and does not necessarily follow the order in the Jacob's ladder \cite{Perdew01AIPCP} (e.g.,\ LDA is more accurate than PBE and even B3LYP). Consistent with previous findings by Hait and Head-Gordon \cite{Hait20JCTC}, RO\wbxv{} gives the best performance here, which is slightly better but otherwise similar to ROMP2.

    \begin{table}[h]
        \centering
        \caption{Excitation energy error (unit: eV) and significant transition amplitudes for selected multi-pole excitations of type $1 \to n$ ($n > 1$). For ROHF/MP2, the transition amplitudes are evaluated at the mean-field level using \cref{eq:transition_amplitudes}.}
        \label{tab:eg_1_to_n}
        \begin{tabular}{lcccc}
            \toprule
            \multirow{2}*{State}
                & \multicolumn{2}{c}{EOM-CCSD}
                & \multicolumn{2}{c}{ROHF/MP2} \\
            \cmidrule(lr){2-3} \cmidrule(lr){4-5}
            & Error & Amplitude & Error & Amplitude \\
            \midrule
            \multirow{3}*{\shortstack[l]{Acetylaldehyde \\ $1$ $^1 A''$}}
                & \multirow{3}*{$0.05$}
                & \multirow{3}*{\shortstack[l]{
                    $10a' \to 6a'':$ $0.54$ \\
                    $10a' \to 4a'':$ $0.33$ \\
                    $10a' \to 8a'':$ $0.13$
                    }}
                & \multirow{3}*{$-0.82/0.03$}
                & \multirow{3}*{\shortstack[l]{
                    $10a' \to 6a'':$ $0.72$ \\
                    $10a' \to 4a'':$ $0.42$ \\
                    $10a' \to 8a'':$ $0.18$
                    }} \\
            & & & & \\
            & & & & \\
            \midrule
            \multirow{2}*{\shortstack[l]{Thiophene \\ $2$ $^1 B_2$}}
                & \multirow{2}*{$0.10$}
                & \multirow{2}*{\shortstack[l]{
                    $1a_2 \to 5b_1:$ $0.50$ \\
                    $1a_2 \to 4b_1:$ $-0.45$
                    }}
                & \multirow{2}*{$-0.63/-0.26$}
                & \multirow{2}*{\shortstack[l]{
                    $1a_2 \to 5b_1:$ $0.81$ \\
                    $1a_2 \to 4b_1:$ $-0.55$
                    }} \\
            & & & & \\
            \bottomrule
        \end{tabular}
    \end{table}

    The results above confirm our expectation that perturbative treatments based on simple spin-symmetry-preserved HF references work well for single-CSF excitations. These include both $1 \to 1$ and $1 \to n$ ($n > 1$) excitations as mentioned above. Two examples of the latter are provided in \cref{tab:eg_1_to_n}, from which one can see that ROHF qualitatively captures the significant transition amplitudes of EOM-CCSD and hence serves as a reasonable reference state for perturbation.

    \begin{table}[h]
        \centering
        \caption{Excitation energy error (unit: eV) and significant transition amplitudes for a multi-pole excitation of type $m \to n$ ($m,n > 1$). For ROHF/MP2, the transition amplitudes are evaluated at the mean-field level using \cref{eq:transition_amplitudes}.}
        \label{tab:eg_m_to_n}
        \begin{tabular}{lcccc}
            \toprule
            \multirow{2}*{State}
                & \multicolumn{2}{c}{EOM-CCSD} & \multicolumn{2}{c}{ROHF/MP2} \\
            \cmidrule(lr){2-3} \cmidrule(lr){4-5}
            & Error & Amplitude & Error & Amplitude \\
            \midrule
            \multirow{6}*{\shortstack[l]{Pyrrole \\ $1$ $^1B_1$}}
                & \multirow{6}*{$0.05$}
                & \multirow{6}*{\shortstack[c]{
                    $2b_1 \to 10a_1:$ $-0.44$ \\
                    $2b_1 \to 14a_1:$ $0.18$ \\
                    $2b_1 \to 16a_1:$ $0.14$ \\[3pt]
                    $1a_2 \to 7b_2:$ $-0.41$ \\
                    $1a_2 \to 10b_2:$ $0.18$
                    }}
                & \multirow{3}*{$-0.85/-0.08$}
                & \multirow{3}*{\shortstack[c]{
                    $2b_1 \to 10a_1:$ $-0.88$ \\
                    $2b_1 \to 14a_1:$ $0.33$ \\
                    $2b_1 \to 16a_1:$ $0.14$
                    }} \\
            & & & & \\
            & & & & \\
            \cmidrule(lr){4-5}
            & & & \multirow{3}*{$-0.34/-0.22$}
                & \multirow{3}*{\shortstack[c]{
                    $1a_2 \to 7b_2:$ $-0.92$ \\
                    $1a_2 \to 10b_2:$ $0.30$ \\
                    $1a_2 \to 9b_2:$ $-0.11$
                    }} \\
            & & & & \\
            & & & & \\
            \bottomrule
        \end{tabular}
    \end{table}

    For the more general $m \to n$ ($m,n > 1$) excitations, the single-CSF ansatz of ROHF becomes insufficient. One such example is provided in \cref{tab:eg_m_to_n}.
    For the $1$ $^1B_1$ state of pyrrole, EOM-CCSD predicts it to be a coherent mixture of two $1 \to n$ transitions, one from $2b_1$ to the $a_1$ manifold and the other from $1a_2$ to the $b_2$ manifold. However, ROHF (and all other HF as well as KS-DFT methods in \cref{tab:stat_PT2,tab:stat_DFT}) can only find them separately, thus missing the important coherence information between the two states, which can affect the computation of non-energetic properties.
    We note that these $m \to n$ excitations will probably become more important for larger systems where an excitation can coherently involve multiple single-CSF transitions from different local motifs of the system that lie close to each other. \cite{Smith10CR,Trinh17SA}
    In such cases, a straightforward solution is to adopt a NOCI-MP2-like strategy in either the perturb-then-diagonalize\cite{Yost13JCP,Yost16JCP} or the diagonalize-then-perturb\cite{Burton20arXiv} manner. This scheme will be explored in future works.

    After benchmarking the performance of PT2 corrections to simple self-consistent excited states, we illustrate the importance of \textit{ab initio} methods using heptazine as an example. Recent computational studies have suggested a small but \emph{negative} singlet-triplet gap (defined as $\Delta E_{\tr{ST}} = E_{\tr{S}_1} - E_{\tr{T}_1}$) for heptazine, \cite{deSilva19JPCL,Ehrmaier19JPCA} which contrasts with the more common scenario where the HOMO-LUMO exchange integral stabilizes $\tr{T}_1$ over $\tr{S}_1$.
    These computational predictions have been supported experimentally by the observed long lifetime of the $\tr{S}_1$ state of a heptazine derivative in the presence of heavy atoms and triplet quenchers. \cite{Ehrmaier19JPCA} Remarkably, it has been shown that only electron correlation methods that include explicitly the effect of double excitations give the correct negative $\Delta E_{\tr{ST}}$, while methods such as CIS and adiabatic TDDFT \cite{Petersilka96PRL,Dreuw05CR} predict a positive gap and hence are qualitatively incorrect. \cite{deSilva19JPCL,Ehrmaier19JPCA}

    \begin{table*}[h]
        \centering
        \caption{Excitation energies of heptazine $\tr{S}_1$ and $\tr{T}_1$ states and the singlet-triplet gap $\Delta E_{\tr{ST}}$ (unit: eV) computed using the various PT2 and KS-DFT methods discussed above. The $\ex{S^2}$ values are listed for HF and KS-DFT. EOM-CCSD and CASPT2 results taken from ref.\ \citenum{Ehrmaier19JPCA} are included for comparison. The cc-pVDZ basis \cite{Dunning89JCP} is used in accordance with ref.\ \citenum{Ehrmaier19JPCA}.}
        \label{tab:heptazine}
        \begin{tabular}{llccccc}
            \toprule
            \multicolumn{2}{l}{\multirow{2}*{Method}}
                & \multicolumn{2}{c}{$\tr{S}_1$}
                & \multicolumn{2}{c}{$\tr{T}_1$}
                & \multirow{2}*{$\Delta E_{\tr{ST}}$} \\
            \cmidrule(lr){3-4} \cmidrule(lr){5-6}
            & & $E$ & $\ex{S^2}$ & $E$ & $\ex{S^2}$ & \\
            \midrule
            \multirow{3}*{U}
                & HF
                    & $1.17$ & $2.19$
                    & $3.61$ & $2.26$
                    & $-2.44$ \\
                & MP2
                    & $5.10$ &
                    & $2.93$ &
                    & $2.17$ \\
                & \wbxv
                    & $2.72$ & $1.22$
                    & $3.20$ & $2.02$
                    & $-0.48$ \\
            \midrule
            \multirow{3}*{RO}
                & HF
                    & $4.24$ & $0$
                    & $4.03$ & $2$
                    & $0.21$ \\
                & MP2
                    & $1.16$ &
                    & $2.20$ &
                    & $-1.04$ \\
                & \wbxv
                    & $3.38$ & $0$
                    & $3.27$ & $2$
                    & $0.11$ \\
            \midrule
            \multirow{2}*{HP}
                & HF
                    & $2.01$ & $1.96$
                    & $1.98$ & $2.42$
                    & $0.03$ \\
                & MP2
                    & $4.99$ &
                    & $4.89$ &
                    & $0.10$ \\
            \midrule
            \multirow{2}*{SP}
                & HF
                    & $0.91$ & $4\times 10^{-5}$
                    & $2.74$ & $2$
                    & $-1.83$ \\
                & MP2
                    & $4.04$ &
                    & $5.37$ &
                    & $-1.33$ \\
            \midrule
            \multicolumn{2}{l}{EOM-CCSD}
                & $2.78$ & & $2.96$ & & $-0.18$ \\
            \midrule
            \multicolumn{2}{l}{CASPT2 (12,12)}
                & $2.33$ & & $2.55$ & & $-0.22$ \\
            \bottomrule
        \end{tabular}
    \end{table*}

    We compute the excitation energies of $\tr{S}_1$ and $\tr{T}_1$ as well as the S-T gap of heptazine using the PT2 variants above and KS-DFT with the \wbxv{} functional. The results are listed in \cref{tab:heptazine}. As can be seen, all SCF methods fail qualitatively except for UHF/KS and SPHF, which both predict a negative S-T gap, but UHF/KS stabilizes $\tr{S}_1$ over $\tr{T}_1$ due to the significant spin-contamination of the $\tr{S}_1$ state (especially UHF) and hence is right for the wrong reason. Indeed, the spin-pure ROHF/KS predicts $\Delta E_{\tr{ST}} > 0$. At the PT2 level, only ROMP2 and SPMP2 improve upon their HF references and give qualitatively correct results. This should be contrasted with the anti-correction brought by UMP2 and HPMP2, and is consistent with the spin-symmetry preservation/breaking of the reference states.
    Despite being qualitatively correct, however, achieving quantitative accuracy is still difficult at the self-consistent PT2 level: both ROMP2 and SPMP2 overestimate the stability of $\tr{S}_1$ compared to EOM-CCSD and CASPT2. This can be attributed to the multi-CSF nature of the heptazine $\tr{S}_1$ state, as already evident from the significant spin-contamination of the UHF wave function, which is not cured by HPHF.


    To conclude, ROMP2 provides accurate state-specific excitation energies for states dominated by a single CSF. The computational cost is modest and scales as non-iterative $O(N^5)$. Moving forward, a production-level implementation is straightforward because it uses only existing ground state code. For example, density fitting \cite{Feyereisen93CPL,Dunlap00PCCP,Jung05PNAS} can be added to reduce the computational cost for no extra work; likewise, analytical gradients can also be incorporated to enable geometry optimization and molecular dynamics for excited states.
    In the future, one can extend ROMP2 to better describe multi-CSF excitations by combining it with NOCI-MP2 \cite{Yost13JCP,Yost16JCP,Yost18JCTC,Burton20arXiv} or
    going higher in the perturbation level (e.g.,\ MP2.5 \cite{Pitovnak09CPC,Riley12PCCP,Sedlak13CPC,Bozkaya14JCP,Bertels19JPCL}).
    With these potential developments, ROMP2 can be used to study photochemical systems such as obtaining accurate Stoke's shifts \cite{Kowalczyk13JCP} and thermally activated delayed fluorescence \cite{Parker61TFS,Tao14AM} (TADF) rates of some dyes.

    \textbf{Supporting Information} (i) The structure files of all molecules in the test set. (ii) The full data and the error distribution plots of the statistics in \cref{tab:stat_PT2,tab:stat_DFT}. (iii) Errors of CIS(D), EOM-CC2, and ADC(2) excitation energies on the test set used in this work.

    \begin{acknowledgement}
        HY thanks Diptarka Hait for useful discussion and proofreading the manuscript. This work was funded by a grant from the NSF (Grant No.\ CHE-1900358).
    \end{acknowledgement}

    \bibliography{refs}

\begin{mcitethebibliography}{111}
\bibitem[ROH()]{ROHF_note}
Some authors prefer to call it opposite-spin ROHF (OS-ROHF) to distinguish it
  from the ROHF method for spin-uncompensated systems. We here follow the
  nonmenclature of ROKS and avoid the use of "OS" to avoid potential confusion
  with opposite-spin-scaled MP2 (OSS-MP2).\relax
\mciteBstWouldAddEndPunctfalse
\mciteSetBstMidEndSepPunct{\mcitedefaultmidpunct}
{}{\mcitedefaultseppunct}\relax
\EndOfBibitem
\end{mcitethebibliography}

\end{document}